\documentstyle[12pt]{article}

\newcommand{\lsim}{\mathrel{\lower4pt\hbox{$\sim$}}
\hskip-12.5pt\raise1.6pt\hbox{$<$}\;}

\newcommand{\gsim}{\mathrel{\lower4pt\hbox{$\sim$}}
\hskip-12.5pt\raise1.6pt\hbox{$>$}\;}

\def\downrightarrow{{\mathrel{\lower7.5pt\hbox{$\rightarrow$}}
\hskip-12.2pt\lower4.9pt\hbox{$\vrule height .25truecm$}\;}}

\def\dra#1#2{\mathop{\mathop{\mathrel{\hbox{$#1$}}}\limits_\downrightarrow}
\nolimits_{\hbox{$\ \ #2$}}}

\begin{document}
\begin{titlepage}
%
%
%
%
%
%
%
\begin{flushright}
SLAC-PUB-6716 \\
TECHNION PH-94-13 \\
November 1994\\
T/E
\end{flushright}
\begin{center}
\vfill
\centerline{\large\bf Pure Leptonic Radiative Decays $B^\pm,
D_s\to\ell\nu\gamma$ and}
\centerline{\large\bf the Annihilation Graph}
\bigskip\bigskip
%
%
%
%
%
%
%
%
%

D. Atwood$^{\rm a}$
G. Eilam$^{\rm b}$
and A. Soni$^{\rm c}$ \\
\bigskip
\end{center}

\begin{flushleft}

a) Department of Physics, SLAC, Stanford University, Stanford, CA\ \ 94309, USA
\\
b) Department of Physics, Technion, Haifa, Israel.\\
c) Department of Physics, Brookhaven National Laboratory, Upton, NY\ \ 11973,
USA\\

\end{flushleft}
\bigskip\bigskip

\bigskip

\vfill

\begin{quote}
{\bf Abstract}:  Pure leptonic radiative decays of heavy-light mesons
are calculated using a very simple non-relativistic model. Dominant
contribution originates from photon emission from light initial quark.
We find $BR(B^\pm\to\ell\nu\gamma)\sim3.5\times10^{-6}$ and $BR(D_s\to
\ell\nu\gamma)\sim1.7 \times10^{-4}$. The importance of these reactions
to clarify the dynamics of the annihilation graph is emphasized.
\end{quote}

\vfill

\begin{center}
Submitted to {\it Physical Review Letters}
\end{center}

\vfill

\hrule
\vspace{5 pt}
\noindent
* Work supported by the Department of Energy Contracts
DE-AC03-76SF00515 (SLAC) and DE-AC02-76CH0016 (BNL).
This work was also supported in part by the United
States-Israel Binational Science Foundation under Research Grant
Agreement 90-00483/3, by the German-Israeli Foundation of Scientific
Research and Development, and the
fund for promotion of research at
the Technion.
\end{titlepage}
%
%
%

\newpage
An important issue in weak decays of charm and bottom mesons that is not
well understood  quantitatively and needs clarification is the
magnitude of the
annihilation graph. This limits our ability to precisely calculate a
variety of important quantities. For one thing it introduces an element
of uncertainty in the calculation of the hadronic width and consequently
the semi-leptonic branching ratio \cite{jane}. Extraction of the value
of the CKM parameters from the experimental data also becomes
problematic \cite{stone}. Furthermore, predictions of CP violating
asymmetries often involve the annihilation graph \cite{greub}. Many
tests of the Standard Model (SM) via rare decays also gets compromised
\cite{atwood}. For instance, in rare decays such as $B\to\rho\gamma$ or
$B\to K^\ast\gamma$, the annihilation graph provides one source of
long range contributions that need to be quantified before these simple
decays can be reliably used to test the SM or to deduce precisely the
value of the mixing angle [4--6]. The pure leptonic radiative decays
such as:

\begin{eqnarray}
B^\pm & \to & \ell\nu\gamma \label{bpm} \\
D_s & \to & \ell\nu\gamma \label{ds}
\end{eqnarray}

\noindent provide  simple reactions that monitor the annihilation graph.
Recently these reactions have been studied in one model \cite{burdman}.
Given the importance of the annihilation mechanism and the unique
cleanliness of the above reactions as a monitor for it, it is clearly
useful to calculate these in several different bound state models.

We will use a very simple non-relativistic model which has been
previously used for exhibiting the importance of the closely related
reactions \cite{bander,eilam}:

\begin{eqnarray}
D^0 & \to & s \bar{d} g \nonumber \\
B,D & \to & u \bar{d} \gamma. \label{dzero}
\end{eqnarray}

\noindent The simplicity of the model has the advantage that it allows
the rate and the differential spectra for the reactions to be readily
calculated in terms of very few parameters. Thus detailed comparisons
with the experiment can be made. Such comparisons should prove very
valuable for illuminating our understanding of the dynamics of these
reactions.

We begin with the amplitude for the pure leptonic
reaction $B\to\ell\nu_\ell$:

\begin{equation}
M= \frac{1}{4} f_B Tr [\Theta (\not\! p_B+m_B) \gamma_5] \label{meq}
\end{equation}

\noindent where $f_B$ is normalized so that $f_\pi\simeq130$ MeV and

\begin{equation}
\Theta = \sqrt{8} G_F (\bar\ell \gamma_\mu P_L\nu) \gamma^\mu P_L
\label{theta}
\end{equation}

\noindent with $P_L\equiv (1-\gamma_5)/2$. Thus

\begin{equation}
\Gamma(B\to\ell\nu) = \frac{m^3_B}{8\pi} G^2_F f^2_B x_\ell (1-x_\ell)^2
|V_{ub}|^2 \label{gamma}
\end{equation}

\noindent with $x_\ell\equiv m^2_\ell/m^2_B$.

The helicity suppression in eqn.~(\ref{gamma}), characterizing the pure
leptonic decay, can be overcome by the emission of a photon (or gluon for
the corresponding reaction into light quarks) \cite{bander,eilam}.
Amongst the Feynman graphs the most important contribution is the one
arising from the photon emission from the initial light quark. Emission
of the photon from the final fermion line is suppressed by powers of
light fermion masses. Emission of the photon from the initial heavy
quark is smaller (in the amplitude)
by a factor of about $m_u/m_b$ compared to the photon
emission from the light initial quark. Thus, in this model, the
amplitude for reaction (\ref{bpm}) is given approximately by

\begin{equation}
M \simeq  \frac{Q_u e g^2_Wf_B V_{ub}}{8m_u(t+u)m^2_W} \; Tr [
(\not\! p_B \not\!\epsilon^\ast \not\! q\gamma^u P_L)( \bar\ell
\gamma_\mu P_L \nu)] \end{equation}

\noindent where $g^2_W=8G_F m^2_W/\sqrt{2}$.

Thus the differential spectra are given by

\begin{equation}
d\Gamma_{\ell\nu\gamma} = \frac{Q^2_u\alpha G^2_F}{16\pi^2m_B} \;
\left(\frac{f_B}{m_u}\right)^2 |V_{ub}|^2 \frac{t^2s}{(m^2_B-s)^2}\;
dsdt \label{dgamma}
\end{equation}

\noindent where $s=(p_\ell+p_\nu)^2$, $t=(p_\ell+q)^2$. Phase space
integration gives:

\begin{equation}
\Gamma_{\ell\nu\gamma} = \frac{Q^2_u\alpha m^5_B}{288\pi^2} G^2_F
|V_{ub}|^2 f^2_B/m^2_u \label{gammaell}
\end{equation}

Thus, in this simple model, the reaction is basically characterized by the
ratio of the pseudoscalar decay constant ($f_B$) to an effective constituent
 mass parameter for the light quark ($m_u$). Intense efforts are underway
using lattice gauge models to calculate $f_B$ \cite{nothing}.
So far, these calculations have an accuracy of about 20\% and in the
next few years one should be able to pin $f_B$ down to  a precision of
about 10\%. The mass parameter in (\ref{gammaell})  is closely related
to the constituent mass of the light quark. It is important to
understand that this bound state picture makes sense only with
constituent masses. These do not vanish in the limit as the current mass
vanishes but rather they go over to a non-perturbative dimensional
parameter in the theory akin to $\langle \bar uu\rangle$.

To get a feel for the rates involved we will use $f_B=175$ MeV \cite{nothing},
$\frac{V_{ub}}{V_{cb}}=.08$, $V_{cb}=0.04$ \cite{stone}
and $m_u=350$ MeV\null. Then we find:

\begin{equation}
BR(B^\pm\to\ell\nu\gamma) \simeq 3.5\times10^{-6} \label{br}
\end{equation}

\noindent Comparing this radiative decay to the pure leptonic decay, say
$B^\pm\to\mu^+\nu$ we get

\begin{equation}
\frac{BR(B^\pm\to\mu\nu\gamma)}{BR(B^\pm\to\mu\nu)} \simeq 16
\label{brfrac}
\end{equation}

\noindent Of course the increase with respect to the electron mode is
much more pronounced.  It is also useful to compare the lepton ($\mu$, $e$)
arising from the decay sequence

\begin{equation}
B\to \dra{\tau}{\mu\nu\nu,e\nu\nu}\!\!\!\!\!\!\!\!\!\!\!\!\!\!\!
\!\!\!\!\!\!\!\!\!\!\!\nu \label{btotau}
\end{equation}

\noindent with $B^+\to\mu(e)\nu\gamma$. From equation (\ref{gamma})

\begin{equation}
BR(B^\pm\to\tau\nu) \simeq 5.1\times10^{-5} \label{brbpm}
\end{equation}

\noindent Thus, for example,

\begin{equation}
\frac{BR(B^+\to\mu^+\nu\gamma)}{BR( B^+\to\tau^+\nu,
\tau^+\to\mu^+\nu\nu)} \sim 0.4 \label{brfracbr}
\end{equation}

For experimental purposes it is also useful to consider the differential
spectra. Indeed the photon energy spectrum for the annihilation reaction
is well known \cite{bander,eilam}:

\begin{equation}
\frac{dN}{d\lambda_\gamma} = \frac{m_B}{\Gamma}\;\frac{d\Gamma_{\ell
\nu\gamma}}{dE_\gamma} = 24\lambda_\gamma(1-2\lambda_\gamma)
\label{dnfrac}
\end{equation}

\noindent where $\lambda_i=E_i/m_B$. This is clearly very distinct from
the steeply falling
bremsstrahlung photon spectrum. The invariant mass ($t$) of the charged
lepton-photon combination is related directly to the energy carried by
the $\nu$ and this along with the lepton energy distribution are given
by:

\begin{eqnarray}
\frac{dN}{d\lambda_\nu} & = & 36(1-2\lambda_\nu) [ 2\lambda_\nu +
(1-2\lambda_\nu) \ell n (1-2\lambda_\nu)] \label{dnfractwo} \\
\frac{dN}{d\lambda_\ell} & = & 36[2\lambda_\ell
(3-5\lambda_\ell)+(1-2\lambda_\ell)(3-2\lambda_\ell) \ell
n(1-2\lambda_\ell)] \label{dnfracthree}
\end{eqnarray}

\noindent These normalized spectra are displayed in Fig.~2.

We can apply this formalism directly to the $D_s$ case, i.e.\ for
$D_s\to\mu\nu\gamma$ or $e\nu\gamma$. The decay constant is already
determined quite well with an accuracy of $\lsim15\%$ to be $f_{D_s} =
230$ MeV \cite{nothing}. Note also that the dominant contribution is now
the emission of the photon
from the strange quark and use of the simple non-relativistic picture
should work better; we use $m_s=500$ MeV and get

\begin{equation}
BR(D_s\to\ell\nu\gamma)\simeq 1.7\times10^{-4} \label{addone}
\end{equation}

\noindent and

\begin{equation}
\frac{BR(D_s\to\mu\nu\gamma)}{BR(D_s\to\mu\nu)} \simeq 3.9\times 10^{-2}
\label{addtwo} \end{equation}
The spectra for $D_s\rightarrow \ell\nu\gamma$ may also
be obtained from
Figure 2 except the role of $\ell$ and $\nu$ are
interchanged from the case for B decays.

If we consider inclusively the final state
$e$ or
$\mu + 0\  hadrons$
in the decay of $B^\pm$
then $\ell\nu\gamma$
dominates over $\ell\nu$
though the later may be distinguished by the fixed energy of the
lepton in the $B$ frame. Another important source of this signal is
the decay chain
(\ref{btotau}).
The resultant
energy
spectrum is:

\begin{eqnarray}
\frac{dN}{d\lambda_\ell} (\tau\to\ell)
&=&
{1+2x_\tau\over 3 x_\tau^2}f(\lambda_\ell)
\theta(x_\tau-2\lambda_\ell)\nonumber\\
&+&
{3-2x_\tau\over 3 (1-x_\tau)^2}f({1\over 2}-\lambda_\ell)
\theta(2\lambda_\ell-x_\tau)
\label{dnfracfour}
\end{eqnarray}
where
$x_\tau=m_\tau^2/m_B$, $\theta$ is the Heavyside
function
and
\begin{equation}
f(\lambda)=16(3-4\lambda)\lambda^2
\label{fdef}
\end{equation}
\noindent This is displayed in Fig.~3 for $B^+$ and also for $D^+_s$.
It is clear from this figure that the
lepton spectrum from
$B \rightarrow\ell\nu\gamma$ is much harder than
the spectrum from (\ref{btotau})
and  may be experimentally distinguished in this way.

In the case of the decay of $D_s$ the corresponding normalized spectra
are also shown in Fig. 3. From (\ref{gamma})
\begin{equation}
BR(D_s\rightarrow \tau \nu)\sim 4.3\%.
\label{brdstau}
\end{equation}
It follows that
\begin{equation}
\frac{BR(D_s\to\mu^+\nu\gamma)}{BR( D_s\to\tau^+\nu,
\tau^+\to\mu^+\nu\nu)} \sim 0.023 \label{bsfracbr}.
\end{equation}
Thus, since the normalized spectra for the charged lepton,
in Fig.3, have similar
shapes for the $D_s$ case, detection of the photon in
$D_s\to\mu^+\nu\gamma$
would be necessary to observe it against
the chain
(\ref{btotau}) for $D_s$ decay.

As more data becomes available it will be very instructive to see how
well this simple picture, outlined above for the important
$\ell\nu\gamma$ modes, works and especially if the data can be accounted
for with $m_u$ of about 350 MeV and $m_s$ of about 500 MeV\null. Let us
briefly contrast this with the model of Ref.~(7).  That model certainly
has interesting features of the heavy quark symmetry \cite{isgur}
built into it. However it also necessitates the introduction of  several
hadronic parameters which  are not readily accessible, at least at
present. Therefore it is difficult  to quantify the  rates and the
detailed spectra. In particular it is useful to note that one of the form
factors of the model of Ref.~(7) also goes as $1/m_u$, characteristic
of the photon emission from the light quark \cite{ambudson}. Needless to
say given the importance of the reactions it would be very helpful if
comparison of the data with both of these as well as other models
\cite{greub2} is pursued vigorously.

G.E. wishes to thank members of the BNL theory group for their
hospitality. We thank Hubert Simma and Daniel Wyler for useful
conversations. This work was supported in part by the United
States-Israel Binational Science Foundation under Research Grant
Agreement 90-00483/3, by the German-Israeli Foundation of Scientific
Research and Development, by an SSC fellowship and DOE contract
DE-AC03-76SF00515 and by DOE contract DE-AC02-76CH00016. Partial support
(for G.E.) was also obtained by the Fund for Promotion of Research at
the Technion.

\eject

\noindent{\Large\bf List of Figures}
\medskip

\begin{enumerate}

\item Feynman graph for the dominant contribution to $B\to\ell\nu\gamma$.
\protect

\item $B\to \ell^- \bar\nu \gamma $ normalized energy spectra are shown. Solid
line is for the photon energy, the dashed is for the neutrino energy
(which is directly related to invariant mass of the electron-photon
combination) and the dash-dot for the electron energy.
For the case of
$D_s\to \ell^+ \nu \gamma$
the dashed curve represents the neutrino energy spectrum while the
dash-dot curve represents the lepton energy since in this case the
roles of the lepton and neutrino are reversed.

\item Comparison of the charged lepton ($\ell $) spectra for $B(D_s)\to
\tau \nu \to 3\nu +\ell $ with $B(D_s)\to \ell \nu \gamma $. Solid line
is for $B$; and dotted one for $D_s$, for the first reaction. The dash
and the dash-dot are for the second reaction for $D_s$ and $B$
respectively.

\end{enumerate}

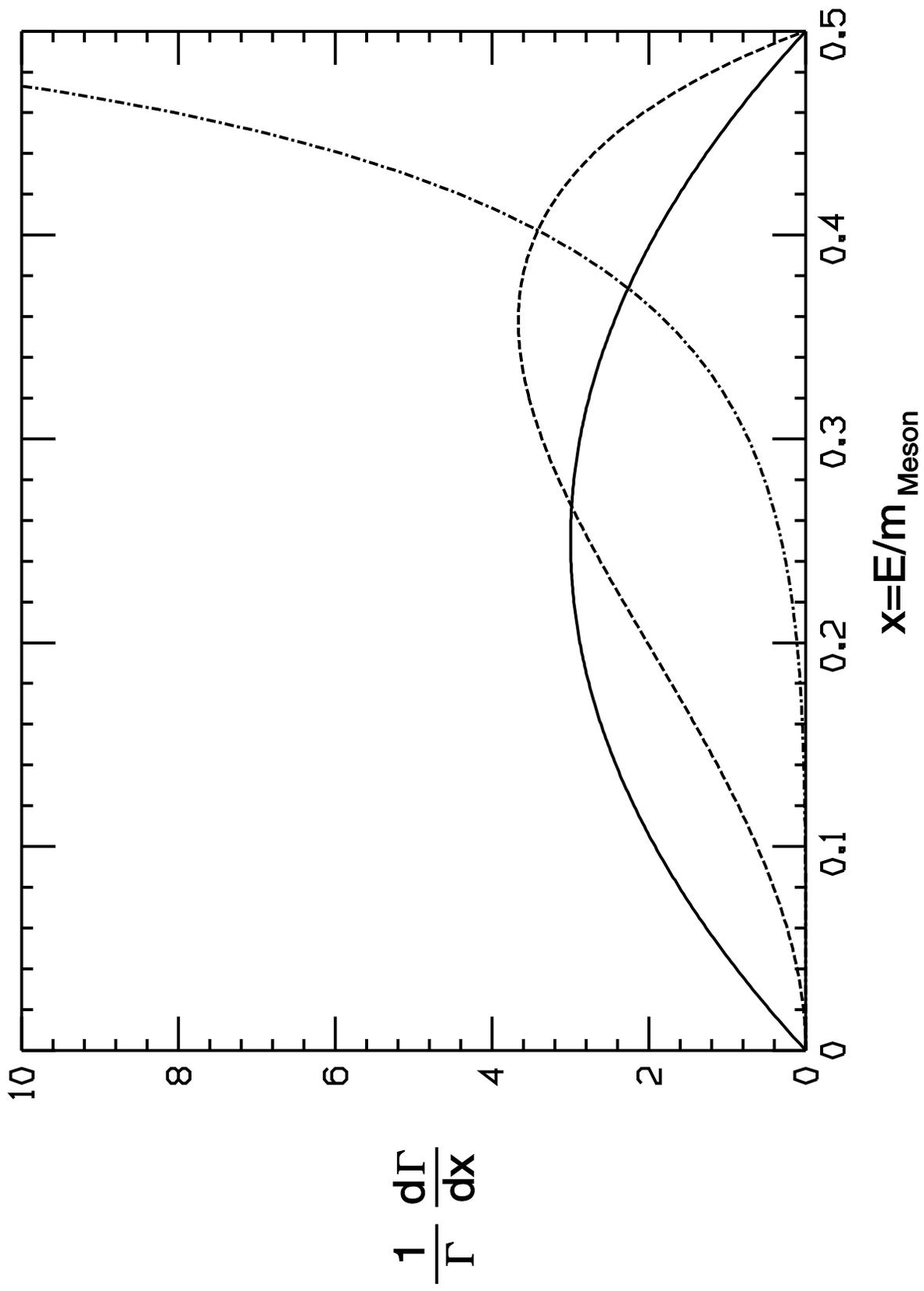
\end{document}